\documentclass{rmaa}
\usepackage{paralist}
\usepackage[latin1]{inputenc}

\title{Ultra-High-Energy Cosmic Rays from a Magnetized Strange Star\\
       Central Engine for Gamma-Ray Bursts}
\author{
  O. Esquivel,\altaffilmark{1,2}
  and D. Page,\altaffilmark{3}}
\altaffiltext{1}{Astronomisches Rechen-Institut am Zentrum f\"ur Astronomie der 
Universit\"at Heidelberg, Heidelberg, Germany}
\altaffiltext{2}{Fellow of IMPRS for Astronomy and Cosmic Physics, Heidelberg.}
\altaffiltext{3}{Departamento de Astrof\'{\i}sica Te\'orica, 
Instituto de Astronom\'{\i}a, UNAM, M\'exico,}
\shortauthor{Esquivel, \& Page}
\shorttitle{UHECRs from a Magnetized SS Central Engine for GRBs}

\fulladdresses{
\item Oscar Esquivel: Astronomisches Rechen-Institut am Zentrum f\"ur Astronomie der 
Universit\"at Heidelberg, M\"onchhofstra{\ss}e 12-14, 69120 Heidelberg, Germany
 (esquivel@ari.uni-heidelberg.de).
\item Dany Page: Departamento de Astrof\'{\i}sica Te\'orica, 
Instituto de Astronom\'{\i}a, Universidad Nacional Aut\'onoma de M\'exico, 
M\'exico D.F. 04510, M\'exico
  (page@astroscu.unam.mx).}
\listofauthors{O. Esquivel, \& D. Page}
\indexauthor{Esquivel, O.}
\indexauthor{Page, D.}

\abstract{Ultra-high-energy cosmic rays (UHECRs) have been tried to be related 
to the  most varied and powerful sources known in the universe. 
Gamma-ray bursts (GRBs) are natural candidates. 
Here, we argue that cosmic rays can be accelerated by large 
amplitude electromagnetic waves (LAEMWs) when the MHD approximation 
of the field in the wind generated by the GRB's magnetized central 
engine breaks down. 
The central engine considered here is a strange star born with differential rotation
from the accretion induced conversion of a neutron star into a strange star in a 
low-mass X-ray binary system.
The LAEMWs generated this way accelerate light ions to the 
highest energies $E = q\eta\Delta\Phi_{max}$ with an efficiency $\eta \sim 10^{-1}$ 
that accounts for all plausible energy losses. 
Alternatively, we also consider the possibility that, once formed, the LAEMWs 
are unstable to creation of a relativistically strong  electromagnetic
turbulence due to an overturn instability. 
Under this assumption, a lower limit to the efficiency of
acceleration is estimated to be about $\eta \sim 10^{-2.5}$. 
Due to their age, low mass X-ray binary systems can be located in regions of low 
interstellar medium density as, e.g., globular clusters or even intergalactic medium
in case of high proper motion systems,
and cosmic ray energy losses due to proton collisions with photons at the decelerating 
region are avoided, thus opening the possibility for particles to exploit the 
full voltage available, well beyond that currently observed.}

\resumen{Se ha tratado de asociar a los rayos c\'osmicos de ultra alta energ\'\i{}a (RCUAE)
a las fuentes m\'as variadas y poderosas que se conocen en el universo.
Los estallidos de rayos gamma (ERG) son candidatos naturales.
En este trabajo sostenemos que tales rayos c\'osmicos pueden ser acelerados por ondas electrom\'agneticas
de alta amplitud (OEAA), las cuales son creadas al momento en que la aproximaci\'on MHD del campo en el viento
generado por el generador central del ERG deja de ser valida.
En tal situaci\'on se considera como generador central a una estrella extra\~{n}a con rotaci\'on diferencial
la cual obtiene la estrella al momento en que una estrella de neutrones se transforma en una estrella extra\~{n}a 
dentro de un sistema binario de rayos X de baja masa. 
Las OEAA generadas de esta manera aceleran los iones ligeros a las m\'as altas energ\'\i{}as, $E = q\eta\Delta\Phi_{max}$,
en donde la eficiencia $\eta \sim 10^{-1}$ da cuenta de todas las posibles perdidas energ\'eticas.
De manera alternativa, tambi\'en consideramos la posibilidad de que una vez formadas, las OEAA se vuelquen entre s\'\i{},
dando paso as\'\i{} a una turbulencia electromagn\'etica relativisticamente poderosa.
Bajo \'esta suposici\'on, se estima un l\'\i{}mite inferior para la eficiencia, $\eta \sim 10^{-2.5}$.
Debido a su edad, los sistemas binarios de rayos X de baja masa pueden localizarse en regiones de baja densidad de
materia interestelar, p. ej., en c\'umulos globulares o inclusive en el medio intergal\'actico en el caso de los sistemas
con un gran movimiento propio; en cuyo caso las perdidas energ\'eticas sufridas por los rayos c\'osmicos en el medio
circundante son evitadas y dichas part\'\i{}culas pueden experimentar el voltage m\'aximo disponible, produciendo
 eventos que rebasar\'\i{}an inclusive las observaciones existentes.}
\addkeyword{Cosmic Rays}
\addkeyword{Gamma-ray sources}
\addkeyword{gamma-ray bursts}
\addkeyword{Neutron stars}

\begin{document}
 
\maketitle 

\section{Introduction}
\label{sec:intro}
A long standing puzzle in astrophysics remains the origin of the highest 
energy cosmic ray events with energies in excess of $10^{20}$ eV 
\citep{2000RvMP...72..689N}. 
The ultra-high-energy cosmic rays (UHECRs) observed so far follow a well 
known energy spectrum $N(E) \sim E^{-2.7}$ beyond the feature known as 
the 'ankle' located at $10^{18.5}$ eV \citep{1994ApJS...90..937A,1994ApJ...424..491B}.
The origin of the particles with energies greater than $\sim 10^{19}$ eV 
is usually attributed to extragalactic sources \citep{1994ApJS...90..937A}.
The distance an UHECR can travel is limited by the energy losses 
due to photo-pion production in collisions with cosmic 
microwave background photons. 
This sets a radius, $R_{GZK} \sim$ 50 Mpc, beyond which the energy losses 
are copious and where the energy of a cosmic ray can be lowered even by a 
few orders of magnitude. 
This mechanism would generate a signature in the spectrum at 
$\sim 5\ \times\ 10^{19}$ eV, known as the GZK cutoff 
\citep{1966PhRvL..16..748G,1966JETPL...4...78Z}.\\
However, in recent years serious constraints on its existence prompted the attention to models containing new physics as the possible solution for the UHECR enigma (see \citealp{arXiv:0801.3028} for a review of the models).

A plethora of astrophysical objects have been considered as possible 
sources of UHECRs.
Among them gamma ray bursts (GRBs) have received special 
attention \citep{1995ApJ...453..883V,1995ApJ...449L..37M,1995PhRvL..75..386W,2002PhRvL..89p1101C,2003ApJ...592..378V}.
In fact, the energetics associated to local GRBs is found to enable a 
GRB-UHECR association \citep{2004APh....21..125W}.
\citet{2004APh....21..125W} considered that a 
good estimate for the local emissivity 
$\dot{\epsilon}_{\scriptscriptstyle{UHECR}}^{\scriptscriptstyle{(GZK)}}$ 
needed to power cosmic rays with energies $\geq 10^{20}$ should be 
$\dot{\epsilon}_{\scriptscriptstyle{UHECR}}^{\scriptscriptstyle{(GZK)}} \approx 
U/t_{\gamma \pi} \approx 10^{44}U_{21}$ ergs Mpc$^{-3}$ yr$^{-1}$, 
where $U=10^{21}U_{21}$ ergs cm$^{-3}$ is the observed energy density in 
$>10^{20}$ eV cosmic rays 
($U_{21} \approx$ 0.5 for the HiRes experiment and $U_{21} \approx$ 2 for AGASA) 
and the timescale of a proton with energy $\sim 10^{20}$ eV for photo-pion 
production is $t_{\gamma \pi}$. 
For a local distribution of GRBs (z\ $\ll$\ 1), which is inside the GZK cutoff, 
pre-BeppoSAX estimates of the luminosity density \citep{1995ApJ...453..883V,1995PhRvL..75..386W} argue for a value of 
$\dot{\epsilon}_{\scriptscriptstyle{GRB}}^{\scriptscriptstyle{(GZK)}} \approx 
10^{44} \dot{\epsilon}_{44}$ ergs Mpc$^{-3}$ yr$^{-1}$, 
where $\dot{\epsilon}_{44} \approx$ few. 
In a more recent analysis, \citet{2003ApJ...592..378V} have arrived at a value of 
$\dot{\epsilon}_{44} \approx$ 1.1.  
Even considering the uncertainties in these estimates, the similarity of the two
luminosity densities 
$\dot{\epsilon}_{\scriptscriptstyle{UHECR}}^{\scriptscriptstyle{(GZK)}}$ 
and $\dot{\epsilon}_{\scriptscriptstyle{GRB}}^{\scriptscriptstyle{(GZK)}}$
make it reasonable to consider that UHECRs may be powered by the local GRBs.

The $\sim 10^ {50} - 10^{52}$ ergs \citep{2004NuPhS.132..255F} required to fit the estimated GRBs' energetics,
 as well as the short observed GRB durations,
naturally point toward gravitational energy associated to compact objects 
as the original energy source.
One broad class of central engine (CE) models consider the merger of two neutron stars 
\citep{1986ApJ...308L..43P}, or a black-hole and a neutron star \citep{1991AcA....41..257P}, in a binary system or
an isolated ``failed'' supernova \citep{1993ApJ...405..273W} in which the released gravitational energy
is temporarily stored in thermal energy and part of it, through neutrinos, is
inefficiently converted into relativistic outflow and successive shocks required 
to fit the GRBs' light curves.
Most of these models eventually involve a black-hole which also swallows a very large
part of the energy.
And in addition to that, the recent analysis of \citep{2006Natur.442.1014S} shows that the supernova-GRB connection
is not as strong as previously tought, inducing \citep{2006Natur.444.1053G} to suggest a new explosive process to explain
the lack of GRB-associated supernova detections.
Another class of CE models takes advantage of the possibility of storing a significant
part of the released gravitational energy into rotational kinetic energy of the 
newly formed compact object \citep{1992Natur.357..472U}, hence limiting the tremendous energy drain
by thermal neutrinos and avoiding the black-hole gluttony.
This kinetic energy can then be extracted on time scales of seconds using the torque 
exerted by the strong magnetic field of the compact star.
In the ``DROCO'' (Differentially Rotating Compact Object) scenario \citep{1998ApJ...508L.113K}
differential rotation of the compact object acts as a powerful dynamo and results in
magnetic field strengths of the order of $10^{17}$ G or higher where subsequent 
buoyancy forces induce emergence of flux ropes at the stellar surface and produce
successive sub-bursts.

\section{Diferentially rotating compact obejcts}
Here, we explore the possibility that UHECRs are accelerated in the Poynting-dominated
outflow of a GRB when the central engine is a new born strange star with a rotation period
$\Omega \sim 10^4$ s$^{-1}$ and a magnetic field of the order of $10^{17}$ G.
Strange stars \citep{1986ApJ...310..261A,1986A&A...160..121H}
are self-bound compact object made entirely of deconfined quark matter,
comprising the three flavors of light quarks $u$, $d$ and $s$,
which is {\em assumed} to be the true ground state of hadronic matter.
Within this ``strange matter hypothesis'' \citep{1971PhRvD...4.1601B,1984PhRvD..30..272W} once a seed of quark matter
appears within a compact star it will rapidly convert the whole star into a 
strange star in a time scale of the order of milliseconds up to a few seconds 
\citep{1994PhRvD..50.6100L,2004ApJ...608..945H}.
If this can happen in core collapse supernovae then all objects we call 
``neutron stars'' may actually be strange stars; 
there is however some evidence that this is not the case
(see, e.g., \citep{2005PrPNP..54..193W} for a discussion).
We will consequently consider a more plausible scenario assuming that the critical 
density for appearance of quark matter is large enough that only the most massive 
neutron stars may convert into strange stars. 
Such a conversion is thus expected to occur in neutron stars which have accreted a 
significant fraction of a solar mass in a low-mass X-ray binary system (LXMB).
The energetics, comparable to a typical core-collapse supernova, 
and the expectable frequency of such events were considered by 
\citet{1996PhRvL..77.1210C} who proposed them as candidates for GRBs.
Given the accretion spin-up of the neutron star and its subsequent shrinking during its
conversion to a strange star, the new-born strange star is plausibly rotating at sub-millisecond 
period while the differential rotation due to the very different density profile of a strange star 
compared to a neutron star can be expected to results in a DROCO as described in \citep{1998PhRvL..81.4301D},
i.e., with a magnetic field of the order of $10^{17}$ G.
An attractive aspect of the strange star scenario is that, most baryonic matter being converted
into quark matter, it avoids the problem of baryon contamination of the outflow 
\citep{1991ApJ...375..209H}.
Nevertheless, the outer crust of the neutron star, with a mass of about $10^{-5}\; M_\odot$,
is most probably not converted into strange matter and rather expelled by the neutrino flux and the highly super-Eddington thermal luminosity of the bare strange star \citep{1998PhRvL..80..230U,2001ApJ...550L.179U,2002PhRvL..89m1101P}, as well as by the flux ropes that emerge at the stellar surface \citep{1998ApJ...508L.113K},
providing the protons which can be accelerated at energies well above $10^{20}$ as we will describe later.
\section{The flux} 
The particle flux which one can expect from such event can be estimated by analogy with standard pulsar physics.
In the magnetosphere of the CE
a number of processes take place that enhance the multiplicity of pairs compared to the proton current \citep{1992Natur.357..472U}, besides the pairs produced by the strange star surface.
For an estimate we consider the pulsar's proton current as determined by the Goldreich-Julian current \citep{1969ApJ...157..869G}  
\begin{equation}
\dot{N}_p = f \; \dot{N}_{GJ} \approx f \; \frac{\Omega^2 B_f R^3}{ec} 
\sim 10^{42} \; \Omega_4^2 \; f \; B_{17} \;\; \mathrm{s}^{-1}
\label{eq1}
\end{equation}
where $\Omega_4 = \Omega/10^4$ s$^{-1}$ and $B_{17}=B/10^{17}$ G, 
and have introduced an enhancement factor $f$ which could be larger than 1. 
We can write the particle flux at the Earth, for energies $\geq 10^{20}$ eV, as
\begin{equation}
I(\geq 10^{20} \mathrm{eV}) \sim 
\frac{c}{4\pi} \; \dot{N}_{\scriptscriptstyle{p}} \Delta t \;
    \alpha \nu_{\scriptscriptstyle{GRB}} \, n_{\scriptscriptstyle{G}}  T_{20} 
\label{eq2}
\end{equation}
where $\Delta t$ is the duration of acceleration activity of the source,
$\nu_{\scriptscriptstyle{GRB}}$ the local GRB rate, a fraction $\alpha$ of them being of the type we consider, $n_{\scriptscriptstyle{G}} = 0.02$ Mpc$^{-3}$ the local density of galaxies and $T_{\scriptscriptstyle{20}}$ the lifetime of protons above $10^{20}$ eV.
For energies in excess of $10^{20}$ eV, correspond propagation times $T_{\scriptscriptstyle{20}} \leq 3 \times 10^8$ yr,
however, the vast majority of the sources within the region limited by $R_{GZK}$ lie at $\sim 40-50$ Mpcs from us, thus it is appropiate to use $T_{\scriptscriptstyle{20}} \sim 3 \times 10^8$ yr \citep{1994PhRvD..50.1892A}. 
So far the true birthrate of local GRBs has not been established. 
The enhancement factor which accounts for the collimation of the luminosity has been calculated recently by a fitting procedure to both the observed differential peak flux and redshift distributions at the same time \citep{2004ApJ...611.1033F}, yielding a local GRB rate 
$\nu_{\scriptscriptstyle{GRB}} \sim 5 \times 10^{-5}$ yr$^{-1}$. 
A reasonable estimate (see below) for $\alpha$ is of the order of 10\%.
This gives us, considering a production life-time $\Delta t \sim 10$~s,
\begin{equation}
I(\geq 10^{20} \mathrm{eV}) \sim 2 \times 10^{-20} f  B_{17} \Omega_4^2 \Delta t_{10} \;
      \mathrm{cm^{-2} \ s^{-1} \ sr^{-1}}
\label{eq3}
\end{equation}
which is safely comparable, and even above if $f > 1$, to the observed flux
$I(\geq 10^{20} \mathrm{eV}) \sim 
2.6^{+2.5}_{-0.6}\ \times\ 10^{-20}\ \mathrm{cm^{-2}\ s^{-1}\ sr^{-1}}$
\citep{2000RvMP...72..689N}. 
Notice that the energy in cosmic rays of $10^{20}$ eV, $\dot{N}_{\scriptscriptstyle{p}}(\Delta t) 10^{20} \mathrm{eV}
\sim 10^{51}$ ergs, represents about 1\% of the total initial kinetic energy of the star.
Since not all the kinetic energy lost by the star will necessarily be detectable, in the GRB, this energy may be comparable, or even larger, than the GRB energy.


Acceleration of cosmic rays in GRBs has been considered, as in many other astrophysical 
objects, to be due to the Fermi acceleration process \citep{1949PhRv...75.1169F}, 
this mechanism taking place either in the plasma outflow \citep{1995PhRvL..75..386W} or
in the external shock of the outflow when the latter hits the interstellar medium 
at a large distance from the CE \citep{1995ApJ...453..883V}. 
Nonetheless, it has been shown by Gallant and Achterberg \citep{1999MNRAS.305L...6G} that the 
energy gain by this mechanism in external shocks when starting  from nonrelativistic 
energies is much lower than predicted previously, 
so in a more realistic situation the Fermi acceleration process could hardly 
reproduce the highest energy events so far observed. \citet{2002PhRvL..89p1101C}  
outlined a different scenario in which Alfv\'{e}n shocks traveling through 
the plasma can accelerate light nuclei in a similar fashion a terrestrial linear 
acceleration does, i.e., the acceleration is stochastic and particles are 
accelerated by the interplay of the  accelerating and decelerating 
phases of the plasma wakefields generated by the Alfv\'{e}n shocks. 
In this stochastic process, the energy losses associated with curvature and 
synchrotron radiation and with ultra-high-energy proton-plasma proton collisions are 
minimal, so the system can be regarded as one in which the acceleration is maximum. 

Instead of considering Alfv\'{e}n shocks, we consider here that large-amplitude 
electromagnetic waves (LAEMWs) are responsible not only of accelerating
particles to produce a GRB event \citep{1994MNRAS.267.1035U} but also responsible for the acceleration of 
plasma protons to the highest energies. 
Poynting flux dominated models of GRBs have been explored recently \citep{2004IJMPA..19.2385Z} as an 
alternative to the fireball model of \citet{1992MNRAS.258P..41R} because
the efficiency of radiation is rather high and the peak photon energy distribution $E_p$ 
could be well reproduced, among other observational facts.
LAEMWs are naturally generated when the MHD approximation of the magnetized 
outflow breaks down. 
This happens at a radius, $R_{MHD}$, from the CE where the plasma density $n_{\pm}$, 
which decreases like $\sim r^{-2}$ \citep{1999ApJ...525..737R}, drops below the density needed for applicability of 
the MHD approximation, the Goldreich-Julian density $N_{GJ}$, which decreases 
like $\sim r^{-1}$ \citep{1969ApJ...157..869G}. 
The LAEMWs generated this way have a frequency $\Omega \sim 10^4$ s$^{-1}$, 
change polarity on a length scale of $2\pi(c/\Omega)\sim 2\ \times\ 10^7$ cm, 
and are very {\em strong} which can be seen if we consider the adimensional 
Lorentz-invariant parameter $\nu \equiv eB/mc\Omega \sim 10^9$ which is the 
ratio of the cyclotron frequency in the wave field $\omega = eB/mc$ to the wave 
frequency \citep{1971ApJ...165..523G}.  
Such a high $\nu$ value can be interpreted by saying that the particles move in a 
nearly static field. Particles {\em surf-ride} \citep{2003ApJ...589..871A} the LAEMWs and suffer a 
ponderomotive force in the radial 
direction due to the inductive electric field orthogonal to the magnetic field 
$f_{pond} = (q/c)\langle \delta v \times \delta B\rangle = mc\Omega \nu$.
Because the particles move relativistically, the radius of curvature associated to
an UHE proton $r_c = 2R_{l}\Gamma^{2}_{w} \gg r$, where $R_l = mc^2\gamma/eB$ is the Larmor radius and $\Gamma_w$ is the wind's Lorentz factor \citep{2003ApJ...589..871A},
and hence excessive synchrotron and adiabatic losses are negligible. 
In principle, following \citet{1975ApJ...196...51R}, the maximum differences in 
the voltage of the wind outflow at large distances from the central engine are
\begin{eqnarray}
\Delta\Phi_{max}  = \frac{\Omega^2 B_f R^3}{2 c^2} \approx 
   1.1 \times 10^{24}\left(\frac{\Omega}{10^4\ \mathrm{s}^{-1}}\right)^2 \nonumber \\
 \times \left(\frac{B_f}{2\times 10^{17}\ \mathrm{G}}\right) 
   \left(\frac{R}{10^6\ \mathrm{cm}}\right)^3\ V.
\label{eq4}
\end{eqnarray}
so that we have an extractable energy well beyond the currently observed data. 
The work produced by the ponderomotive force can be approximated by
\begin{equation}
 W_{pond}\ \approx\ q\eta\Delta\Phi_{max}
\label{eq5}
\end{equation}
where $\eta$ is the efficiency factor.  
From analysis of pulsar wind nebulae it is seen that about 10\% of the total
voltage is transferred to the ions \citep{2003ApJ...589..871A}.

In the situation prevailing in a GRB outflow, an  UHE proton can  loose some of its energy 
due to collisions with ambient protons.
This can be estimated in an easy way \citep{2002PhRvL..89p1101C} by integrating
the collision probability, $\sigma_{pp}n_{p}(r)/\gamma$, from $R_{MHD}$, 
the transition radius, to $R_{MHD} + R_{mfp}$ where $R_{mfp}$
is the mean-free path radius and the collision cross section $\sigma_{pp} \sim 30$ mb. 
If we want the system be collision-free, the threshold condition
\begin{equation}
 \frac{\sigma_{pp}N_{GJ}R_{MHD}}{\Gamma_w}\ \leq\ 1
\label{eq6}
\end{equation}
should be satisfied,
where we have taken $n_{p}(r) = N_{GJ}$ for the plasma density at the transition region. 
It is easy to see that for appropriate values this condition is always satisfied. 

However, as an alternative to this high efficiency acceleration mechanism, we have also 
considered the possibility that 
the interaction of LAEMWs with the plasma can lead to instabilities. 
\citet{2001MNRAS.321..177L} argued that the interaction of LAEMWs with the plasma 
can lead to violent instabilities where the MHD approximation breaks down 
because an {\em overturn instability} of LAEMWs, creating a broad spectrum of 
random electromagnetic fields.
The mixing of various parts of the wave destroys its oscillatory structure 
thus yielding a relativistically strong electromagnetic turbulence. 
When the correlation length, $l_c$, of the turbulent electric field is much 
shorter than the radiative length, $l_c\ \ll\ l_r$, there will be an effective 
root-mean-squared deficit in the coherence of the ion velocity and the 
accelerating electric field.  
If the turbulent energy is concentrated near the smallest scale $\sim$ skin depth, 
a lower limit to the efficiency of acceleration, $\eta$, is obtained, 
which in case of protons can be
expressed in terms of the classical proton radius $r_p$ and the wave frequency 
$\Omega$, $\eta = (2r_p\Omega/c)^{1/7} \approx 10^{-2.5}$. 
We can see that even if this lower factor is taken into account, 
we have an extractable voltage sufficient to account for the highest energy events. 

Whether the turbulence is considered or not, the principal energy losses for a 
high energy proton once the acceleration process is almost completed are 
due to collisions with the intense thermal photon
field left after adiabatic expansion of the relativistic outflow. 
Each collision undermines $\sim 10\%$ of the UHE proton energy \citep{1994PhRvD..50.1892A} although the 
characteristic time for this energy loss mechanism \citep{1995PhRvL..75..386W} 
$t_{p\gamma} \sim 10/n_{\gamma}\sigma_{p \gamma } c$ is larger than the time of 
expansion of the outflow even when this is entering the decelerating regime.
The collision cross section in this case is 
$\sigma_{p \gamma } \approx 10^{-29}$ cm$^2$.
The photon number density $n_{\gamma}$ was calculated by using the photon
luminosity in the decelerating region,
$L_{\gamma} = 4\pi r^{2}_{d} c \gamma n_{\gamma}\epsilon_{\gamma}$, 
where $\epsilon_{\gamma}$ is the observed photon energy, i.e.,  
$\epsilon_{\gamma} \sim$ 1 MeV.

Due to their age, LMXB systems are not associated with star formation regions and
an interesting possibility arises if we consider the fraction of LXMB systems related 
to globular clusters, $\sim$ 10\% \citep{2001A&A...368.1021L}, in wich the CE
is surrounded by a very tenous ambient medium.
We would thus expect in this case that an equal fraction of the GRBs discussed here would not produce an afterglow, or at least a very faint event \citep{2005A&A...439L..15C}.
This is another feature which avoids energy losses of UHECRs due to the photon field
of the afterglow.
However, we cannot exclude that the impact of the burst on the strange star's companion
will produce a detectable signal \citep{2004MNRAS.349L..38R}, but with negligible effect on
the bulk of UHECRs since it will be of small angular size, restricted to the direction of
the companion, and, in addition, the zone where the acceleration takes place 
is far beyond the location of the companion.

Details of the expected UHECR spectrum are impossible to predict with confidence.
In case of pure electromagnetic spindown of the central engine \citep{2003ApJ...589..871A} 
an $E^{-1}$ dependence is expected
while if acceleration can be regarded as a quasi-stochastic process a spectrum 
$\sim E^{-2}$ would be more appropriate \citep{2002PhRvL..89p1101C}.
Propagation effects are usually argued to cause a steepening of the spectrum, and to that respect it 
has been argued that the consistancy between the data obtained by the HiresI-II, Yakutsk, and Auger
experiments proves the existence of a GZK cutoff in oposition to the results obtained by the Akeno-AGASA experiment,
\citep{arXiv:0801.3028,arXiv:0801.1986}, but the lack of robust data makes impossible to reach a final
conclusion. The model  as presented  here would agree with an extension of events beyond the highest 
energy values observed, $> 10^{20}$ eV.

Finally we mention that the above outlined scenario can be applied equally well
to the accretion induced collapse of a white dwarf into a neutron star,
considering a cataclismic variable as an initial system instead of a low
mass X-ray binary.
However, a young strange star with a bare quark surface will emit, for a possibly long 
period of time, super-Eddington luminosities through pair production at its
surface \citep{1998PhRvL..80..230U,2001ApJ...550L.179U,2002PhRvL..89m1101P}
and later through electron-electron bremsstrahlung from
its electrosphere \citep{2004PhRvD..70b3004J}, with unique spectral characteristics which
would allow to identify it unambiguously.
On the contrary, a neutron star born through accetion induced collapse of a white
dwarf is difficult, if not impossible, to be differenciated from its sibblings
born through the core collapse of a massive star.
Identification of strange stars in binary systems (see, e.g., for some possibilities \citep{2005ApJ...635L.157P})
would give support to our scenario.


\begin{acknowledgments}
The author thanks V. V. Usov for comments on the manuscript.
This work was supported by grants from the Mexican Conacyt (\# 36632-E)
and from UNAM-DGAPA (\# IN112502).
\end{acknowledgments}

 

\end{document}